\documentclass[floats,aps,prl,draft,preprint,showpacs]{revtex4}
\begin{document}
\title{Impurity-induced stabilization of Luttinger liquid in
quasi-one-dimensional conductors}

\author{S.N. Artemenko}

\address{Institute for Radioengineering and Electronics of Russian
Academy of Sciences, 11-7 Mokhovaya str., Moscow 125009, Russia}
\date{\today}
\begin{abstract}
It is shown theoretically that the Luttinger liquid phase in
quasi-one-dimensional conductors can exist in the presence of
impurities in a form of a collection of bounded Luttinger liquids.
The conclusion is based upon the observation by Kane and Fisher
that a local impurity potential in Luttinger liquid acts, at low
energies, as an infinite barrier. This leads to a discrete
spectrum of collective charge and spin density fluctuations, so
that interchain hopping can be considered as a small parameter at
temperatures below the minimum excitation energy of the collective
modes. The results are compared with recent experimental
observation of a Luttinger-liquid-like behavior in thin NbSe$_3$
and TaS$_3$ wires.
\end{abstract}

 \maketitle

Electronic properties of one-dimensional (1D) metals are known to
be very different from those of ordinary three-dimensional (3D)
metals (for a review see Ref.\cite{Voit,FishGlaz,SchulzR}). 3D
electron gas is well described by Landau's Fermi-liquid picture in
which interaction modifies free electrons making them
quasiparticles that behave in many respects like non-interacting
electrons. In contrast to the 3D case, in 1D electronic systems
the Fermi-liquid picture breaks down even in the case of the
arbitrarily weak interaction. In 1D metals, the single-electron
quasiparticles do not exist, and the only low energy excitations
turn out to be charge and spin collective modes with the
sound-like spectrum. These modes are dynamically independent
giving rise to a spin-charge separation in 1D systems.
Furthermore, correlation functions at large distances and times
decay as a power law with interaction dependent exponents. Such a
behavior has been given a generic name Luttinger liquid
\cite{Haldane}.

The concept of Luttinger liquid is of great interest in view of
its application to real physical systems, such as carbon nanotubes
and semiconductor heterostructures with a confining potential
(quantum wires and quantum Hall effect edge states). The case of a
special interest are quasi-1D conductors, \textit{i. e.}, highly
anisotropic 3D conductors with chain-like structure. Numerous
experimental studies of both organic and inorganic q1D conductors
at low temperatures typically demonstrate broken-symmetry states,
like superconductivity, spin- or charge-density wave (CDW) states,
and a metallic behavior above the transition temperature with
non-zero single-particle density of states at Fermi energy. For
instance, the most studied inorganic quasi-1D metals undergo the
Peierls transition from metallic state either to a semiconducting
CDW state (\textit{e. g.}, blue bronze K$_{0.3}$MoO$_3$, TaS$_3$,
(TaSe$_4$)$_2$I \textit{etc}.) or to semimetallic CDW state
(NbSe$_3$) \cite{Gruner}. Typically, this transitions occur in the
temperature range 50 - 250 K. From the theoretical point of view,
the formation of Luttinger liquid in quasi-1D conductors at low
enough temperatures is also problematic because of the instability
towards 3D coupling in the presence of arbitrarily small
interchain hopping \cite{PriFirs,BraYak,Schulz,Boies,Arrig}. So
the interchain hopping induces a crossover to 3D behavior at low
energies, while Luttinger liquid behavior can survive only at high
enough energy scale where it is not affected by 3D coupling.

In contrast to the interchain hopping, the Coulomb interaction
between the electrons at different chains does not destroy the
Luttinger liquid state, the main difference from the 1D case being
the absence of simple scaling relations between the exponents of
the various correlation functions
\cite{Barisic,BoBa,SchulzCoul,MKL}.

However, in recent experimental studies of temperature and field
dependence of conductivity of TaS$_3$ and NbSe$_3$ in
nanoscale-sized crystals a transition from room-temperature
metallic behavior to nonmetallic one accompanied by disappearance
of the CDW state at temperatures below 50 - 100 K was observed
\cite{ZZPM,ZZ,ZZGSZ}. The low temperature non-metallic state was
characterized by power law dependencies of the conductivity on
voltage and temperature like that expected in Luttinger liquid, or
by more strong temperature dependence corresponding to the
variable-range hopping. Resembling dependencies of conductivity
were reported although in focused-ion beam processed or doped
relatively thick NbSe$_3$ crystals \cite{ZZGSZ}. Hopping
conductivity was also detected in dirty quasi-1D conductors KCP
and organic TCNQ-based metals \cite{organics}, while pure
materials are known to undergo the Peierls transition to the CDW
state.

In order to account for such behavior, we study the possibility of
impurity-induced stabilization of a gapless Luttinger liquid state
in quasi-1D metals. Impurities in Luttinger liquid are known to
act as infinite barriers forming the effective boundaries for low
energy excitations \cite{KaneFischer,MatveevGlazman,FuruNag}. This
leads to a dimensional quantization and, consequently, to a
minimal excitation energy $\omega_1$. As a result, the interchain
hopping does not destroy the Luttinger liquid phase at
temperatures $T \ll \omega_1$, producing only small perturbations
of the 1D picture. To show this, we consider first the gapless 1D
Tomonaga-Luttinger (TL) model with impurity potential included,
and make certain that the system with impurities breaks up into a
set of independent segments described as bounded Luttinger liquid
with discrete spectrum. Then we calculate corrections caused by
the interchain hopping to thermodynamic potential and to the
one-particle Green's function, and find that such corrections are
small at low temperatures. Finally, we discuss modifications
introduced by generalization of the TL model to the more realistic
case of Coulomb potential and compare our results with
experimental data.

First of all we start with the TL model ignoring interchain
hopping integral $t_\perp$ in order to formulate the problem in
the zeroth approximation in $t_\perp$. Electronic operators for
right ($r=+1$) and left ($r=-1$) moving electrons with spin $s$
are given in terms of phase fields as (see
Ref.\cite{Voit,SchulzR})
\begin{equation}
\psi_{r,s}(x) = \lim_{\alpha\rightarrow
0}\frac{e^{irk_Fx}}{\sqrt{2\pi \alpha}}\eta_{r,s}e^{iA_{r}},\quad
A_{r} = \frac{1}{\sqrt{2}}[\Theta_\rho-r\Phi_\rho
 + s(\Theta_\sigma - r\Phi_\sigma)] \label{psi}
\end{equation}
here phase fields $\Phi_\nu(x)$ are related to charge ($\nu=\rho$)
and spin ($\nu=\sigma$) densities, while fields $\Theta_\nu(x)$
are related to the momentum operators $\Pi_\nu = (1/\pi)
\partial_x\Theta_\nu$ canonically conjugate to $\Phi_\nu$. Further,
$\eta_{r,s}$ are Majorana ("real") Fermionic operators that assure
proper anticommutation relations between electronic operators with
different spin $s$ and chirality $r$, and the cut off length
$\alpha$ is assumed to be of the order of interatomic distance.

We describe the intrachain properties of the system by the
standard TL Hamiltonian \cite{Voit,SchulzR} with added $2k_F$
impurity backscattering term \cite{FishGlaz}. In the bozonized
form it reads
\begin{eqnarray}
&& H = \sum_{\nu=\rho,\sigma} \int dx \left\{\frac{\pi
v_{\nu}K_\nu}{2} \Pi_\nu^2 + \frac{v_{\nu}}{2\pi K_\nu}
\left(\partial_x \Phi_\nu \right)^2\right\}
\nonumber \\
&&
+ 
\sum_i V_0 d \delta (x-x_i) \cos{(\sqrt{2} \Phi_\rho +
2k_Fx)}\cos{(\sqrt{2} \Phi_\sigma(x))}, \label{H0}
\end{eqnarray}
where $v_{\nu}$ are velocities of the charge ($\nu=\rho$) and spin
($\nu=\sigma$) modes, $K_\nu = v_F/v_\nu$ is the standard
Luttinger liquid parameter describing the strength of the
interaction, $V_0$ and $d \sim \alpha$ are amplitude and radius of
the scattering potential, respectively.

Kane and Fischer \cite{KaneFischer} found that the backscattering
impurity potential for repulsive potential ($K_\rho <1$) flows to
infinity under scaling. Their arguments were generalized by
Fabrizio and Gogolin \cite{FabGogolin} to the case of many
impurities. It was shown that the impurity potential can be
considered as effectively infinite provided that the mean
distance, $l$, between impurities satisfies the condition
\begin{equation}\label{co}
l \gg \frac{1}{k_F} \left(\frac{D}{V_0}\right)^{2/(1-K_\rho)},
\end{equation}
where $D$ is the bandwidth. We assume that the impurity potential
is of atomic scale, $V_0 \lesssim D$, and the interaction between
electrons is not too weak, (\textit{i. e.}, $K_\rho$ is not too
close to 1). Then condition (\ref{co}) is satisfied for $l \gg
1/k_F\sim \alpha$ which is of the order of interatomic distance.
So the limit of strong impurity potential should be a good
approximation in a wide range of impurity concentrations.

Further, $\Pi_\nu$, $\Theta_\nu$ and $\Phi_\nu$ must obey the
commutation relations (see Ref.\cite{Voit,FishGlaz,SchulzR})
ensuring anticommutation of electronic operators (\ref{psi}).
Using then the analogy of Eq.(\ref{H0}) with the Hamiltonian of an
elastic string strongly pinned at impurity sites, we can write
down solutions for the phase operators $\Phi_\nu$ and $\Theta_\nu$
in the region between impurity positions at $x=x_i$ and $x_{i+1}$
as
\begin{eqnarray}
&& \Phi_\nu (x) = \sum_{n=1}^\infty \sqrt\frac{K_\nu}{n}(b_n +
b^+_n) \sin{q_n \tilde x} + \frac{\tilde x\Phi_{i+1} - (\tilde
x-l_i)\Phi_{i}}{\sqrt{2}l_i} \delta_{\nu\rho}  - \sum_{j<i}\pi
\Delta N_{\nu j} - \pi \Delta N_{\nu i} \frac{\tilde x}{l_i},
\nonumber \\
&& \Theta_\nu (x) = \sum_{n=1}^\infty \sqrt\frac{1}{K_\nu n}(b_n
- b^+_n) \cos{q_n \tilde x} + \theta_\nu, \label{Th0}
\end{eqnarray}
where $\tilde x = x-x_{i},$ $l_i=x_{i+1}-x_{i}$, $q_n =\pi n
/l_i$, $\Phi_{i}$ is the modulo $2\pi$ residue of $2k_Fx_i$.
Further, $\Delta N_{\rho i} = (\Delta N_{\uparrow i} +
N_{\downarrow i})/\sqrt{2}$, $\Delta N_{\sigma i} = (\Delta
N_{\uparrow i} - N_{\downarrow i})/\sqrt{2}$, and $\Delta
N_{\uparrow i}$ ($\Delta N_{\downarrow i}$) is the number of extra
electrons with spin up (down) in the region between $i$-th and
$(i+1)$-th impurities, and, finally, $\theta_{\nu i}$ is the phase
canonically conjugate to $\Delta N_{\nu i}$ ($[\theta_{\nu
i},\Delta N_{\nu i}]=i$).

Excitation spectra of the eigenmodes are $\omega_\nu = n
\omega_{1,\nu}\equiv v_\nu q_n$ where $\omega_{1,\nu} = \pi
v_\nu/l_i$ is the minimum excitation frequency for mode $\nu$.

Note that if we consider the open boundary conditions at the
sample ends, $x=0$, and, $x=L$, (instead of periodic boundary
conditions that are commonly used) then operators $\eta_{s}$ in
Eq.(\ref{psi}) are the same for electrons going right and left. In
this case, the electron field operator, $\psi_{s}(x)= \psi_{s+}(x)
+ \psi_{s-}(x)$, vanishes at impurity positions, $x=x_i$, and
expressions for the phase fields between the impurity sites turn
out to be similar to those found for bounded 1D Luttinger liquid
in Refs.\cite{FabGogolin,EggJohMat,Mat}, the main difference being
the summation over $j < i$ that insure proper commutation
relations between the electron operators related to different
segments between impurities. Thus the system breaks up into a set
of independent segments described as bounded Luttinger liquid with
discrete spectrum.

Now we consider the role of interchain hopping adding to
(\ref{H0}) the hopping Hamiltonian
\begin{eqnarray}
&& H_\perp = t_\perp \sum_{m,n,r,s} \int dx \psi^+_{r,s,m}(x)
\psi_{r,s,n}(x) + HC \nonumber \\
&& = \sum_{m,n,r,s} \int dx \frac{it_\perp
\eta_{r,s,n}\eta_{r,s,m}}{\pi \alpha} \left[ \sin(A_{r,m}-A_{r,n})
+ \sin(A_{r,m}-A_{-r,n} + 2irk_Fx) \right],
   \label{Ht}
\end{eqnarray}
where indices $n$ and $m$ denoting the chain numbers related to
the nearest neighbors are added.

Arguments by Schulz \cite{Schulz} on instability of the Luttinger
liquid in the presence of the interchain hopping were based on
calculations of temperature dependence of the thermodynamic
potential at low temperatures. So we calculate contribution of the
interchain hopping to the thermodynamic potential per unit volume
given by the standard expression \cite{AGD}
\begin{equation}\label{ome}
\Delta \Omega = -T \ln \langle S \rangle/V, \quad S=T_\tau \exp
\left(-\int_0^{1/T} H_\perp(\tau) d\tau \right),
\end{equation}
where $V$ is the volume, $T_\tau$ stands for imaginary time
ordering, and $\langle \cdots \rangle$ means thermodynamic
averaging over the unperturbed state.

At temperatures $T \gg \omega_{1,\nu}$, the discreteness of the
excitation spectrum can be neglected, hence, according to Refs.
\cite{PriFirs,BraYak,Schulz,Boies,Arrig}, interchain hopping is
expected to give significant contributions destroying the
Luttinger liquid. We examine the opposite limit, $T \ll
\omega_{1,\nu}$, which does not exist in pure infinite Luttinger
liquid.

Consider first the second order correction in $t_\perp$. The
leading contribution to $\langle S \rangle$ in Eq.(\ref{ome}) is
given by items in which the term related to a given chain contains
contributions from the electrons with the same chirality, $r$,
only,
\begin{equation}
\sum_{m,r,r'} \frac{t_\perp^2}{8\pi^2\alpha^2} \int d{\mathbf 1}
d{\mathbf 2} \langle T_\tau e^{i[A_{r,m}({\mathbf
1})-A_{r',n}({\mathbf 1})]} e^{-i[A_{r,m}({\mathbf
2})-A_{r',n}({\mathbf 2})]} \rangle e^{i(r-r')k_F (x_1-x_2)} ,
\label{exp-op}
\end{equation}
where ${\mathbf 1}=\{x_1,\tau_1\}$ and ${\mathbf
2}=\{x_2,\tau_2\}$. Other items in which the terms related to the
same chain contain contributions from electrons moving both left
and right give small contribution, and we do not discuss them in
details.

Then we use Eq.(\ref{Th0}) in (\ref{psi}) and calculate average in
(\ref{exp-op}) using the relation
\begin{equation}
 \langle T_\tau e^{iA_{r,m}({\mathbf 1})} e^{-iA_{r,m}({\mathbf 2})}
\rangle   = e^{-\frac{1}{2}\langle A_{r,m}^2({\mathbf 1}) +
A_{r,m}^2({\mathbf 2}) - 2T_\tau [A_{r,m}({\mathbf
1})A_{r,m}({\mathbf 2})] \rangle}.\label{exp}
\end{equation}
Neglecting small corrections $\propto
\exp{[-\frac{\omega_{1,\nu}}{T}]}$ due to Planck's distribution
functions, we find for the average in the exponent
\begin{equation}
\langle T_\tau A_{r}({\mathbf 1})A_{r}({\mathbf 2})\rangle =
\frac{1}{8}\ln \left[\frac{(\cosh z - \cos y_+)^{(K_\nu -
K_\nu^{-1})}}{(\cosh z - \cos y_-)^{(K_\nu + K_\nu^{-1})}} \right]
 + i \tan^{-1}\left[ \frac{\sin y_-}{e^{z} - \cos y_-}
\right],\label{aa}
\end{equation}
where $y_{\pm}= \frac{\pi(\tilde x_1 \pm \tilde x_2)}{l_i}$, $z=
\frac{\pi(\alpha + v_\nu |\tau_1-\tau_2|)}{l_i}$ (chain indices
are dropped for brevity here).

In the integrations over ${\mathbf 1}$ and ${\mathbf 2}$, the
leading contributions comes from region ${\mathbf 1} \approx
{\mathbf 2}$, \textit{i. e.}, $|y_-| \ll 1$, $z \ll 1$ where
expression
(\ref{exp-op}) reduces to
\begin{equation}
\frac{t_\perp^2 m L}{\pi^2\alpha^2 T} \int \frac{\cos
[(r-r')k_F(x_-)]\;dx_- d\tau_-}{\prod_{\nu=\rho,\sigma}
[(1+\epsilon_\nu \tau_- )^2 + (x_-/\alpha)^2]^{1+2\delta_\nu} }
\label{ln}
\end{equation}
 where $\epsilon_\nu = \frac{v_\nu}{\pi\alpha}$,
$\delta_\nu = \frac{1}{4}(K_\nu + 1/K_\nu -2)$, $\tau_- =
|\tau_1-\tau_2|$, $x_- = x_1-x_2$, and $m$ is the number of the
nearest-neighbor chains. Contribution to expression (\ref{exp-op})
from integration over region $|y_-| \gtrsim 1$, $z \gtrsim 1$, is
small, $\sim (\alpha/l_i)^{2\delta}$, $\delta = \delta_\rho +
\delta_\sigma$, because $\delta$ is not too small in the assumed
case of the not too small interaction (\textit{cf.} Eq.
(\ref{co})).

Additional items in $\langle S \rangle$ in Eq.(\ref{ome}) in which
the terms related to the same chain contain contributions from
electrons moving both left and right is smaller than that given by
Eq. (\ref{ln}) by factor $\sim
(\alpha/l_i)^{K_\rho + K_\sigma}$.  For reasonable values of
$K_\nu$, this contribution is small and can be neglected.

Similarly, the leading contribution to $\Delta \Omega$ from
higher-order terms in series expansion of the exponential in
Eq.(\ref{ome}) was found to come from even powers $2n$ in
$t_\perp$  that can be represented as a sum of $(2n-1)!!$ items
like (\ref{exp-op}) with almost coinciding times and coordinates.
Therefore, summing up the leading contributions and inserting them
into Eq.(\ref{ome}) we can calculate the variation of the
thermodynamic potential per single chain and per unit length
\begin{equation}\label{omeg}
\Delta \Omega = - a \frac{t_\perp^2 m}{\pi^2 v_F},  a = \int
\frac{(1+\cos 2k_F \alpha x)\;dx d\tau}{\prod_{\nu=\rho,\sigma}[
(1+\tau/K_\nu)^2 + x^2]^{1+2\delta_\nu}}.
\end{equation}
For moderate repulsion, $\delta \sim 1$, $a \sim 1$. In the limit
of strong repulsion, $K_\rho \ll K_\sigma \sim 1$, $a$ is small,
$a \sim K_\rho^2$.

Thus $\Delta \Omega$ is much smaller than the thermodynamic
potential of purely 1D Luttinger liquid, $\Omega_0 \sim
\left(\frac{1}{K_\rho} + \frac{1}{K_\sigma} \right)\varepsilon_F
k_F$,
$$
\Delta \Omega/ \Omega_0 \sim
\left(\frac{t_\perp}{\varepsilon_F}\right)^2,
$$
and temperature-dependent corrections to Eq.(\ref{omeg}) are
determined by small thermally activated contributions  $\propto
\exp{[-\frac{\omega_{1,\nu}}{T}]}$.

Now we calculate modification of the one-particle Green's function
due to the interchain hopping.
\begin{equation}
G({\mathbf 1},{\mathbf 1'}) = -\langle T_\tau \psi ({\mathbf 1})
\bar{\psi} ({\mathbf 1'}) S \rangle / \langle S \rangle. \label{G}
\end{equation}
Again, we consider the low-temperature limit, $T \ll
\omega_{1,\nu}$, non-existing in pure infinite system. Consider
first the second order correction in $t_\perp$ to the Green's
function of pure 1D system, $G_0({\mathbf 1},{\mathbf 1'})$.
\begin{equation}
G_2({\mathbf 1},{\mathbf 1'}) = -\langle T_\tau \psi ({\mathbf 1})
\bar{\psi} ({\mathbf 1'}) S_2 \rangle  + \langle T_\tau \psi
({\mathbf 1}) \bar{\psi} ({\mathbf 1'}) \rangle  \langle S_2
\rangle. \label{G2}
\end{equation}
Calculation is similar to that considered above, (\textit{cf}.
(\ref{exp-op}-\ref{aa})). However, in contrast to the case of the
thermodynamic potential where the leading contribution was given
by regions of almost coinciding values of times and coordinates,
such contributions from two terms in (\ref{aa}) cancel each other.
So the second-order correction is estimated as
$$
G_2({\mathbf 1},{\mathbf 1'}) \lesssim \left(\frac{t_\perp
l}{v_F}\right)^{2} \left(\frac{\alpha}{l}\right)^{2\delta}
G_0({\mathbf 1},{\mathbf 1'}).
$$
Estimation of the fourth order correction in $t_\perp$ gives $G_4
\sim (t_\perp l/v_F)^2(\alpha/l)^{2\delta} G_2$. Therefore, we
conclude that at $T \ll \omega_{1,\nu}$ the interchain hopping
gives small corrections to the one-particle Green's function,
provided
\begin{equation}
\left(\frac{t_\perp l}{v_F}\right)^{2}
\left(\frac{\alpha}{l}\right)^{2\delta} \sim
\left(\frac{t_\perp}{\varepsilon_F}\right)^2
\left(\frac{\alpha}{l}\right)^{2\delta-2} \ll 1, \label{G2e}
\end{equation}
where we estimated the cut-off parameter as $\alpha \sim 1/k_F$.

So far we considered the TL model in which interaction is
described by coupling constants related to forward- and
backscattering. In order to make comparison with experimental data
we must consider a more realistic Coulomb potential. It is
reasonable to assume that the long-range part of the interaction
is dominated by the Coulomb potential, while the backscattering is
described by relatively small coupling constant $g_1$. This
enables us to concentrate on the spin isotropic case and to ignore
the possibility of the spin gap. The problem of the long-range
Coulomb potential on an array of chains was solved in Refs.
\cite{Barisic,BoBa,SchulzCoul}. It was found that interaction of
electrons on a given chain is screened by the electrons on other
chains, and the Coulomb interaction can be described by the TL
Hamiltonian with coupling constants dependent on wave vector,
\begin{equation} \label{g2}
g_2 =g_4 = \frac{4\pi e^2}{s^2(q_\|^2 + \epsilon_\perp
q_\perp^2)},
\end{equation}
where $s$ is the lattice period in the direction perpendicular to
the chains, and $\epsilon_\perp$ is a background dielectric
constant for the transverse direction. Coupling constants in spin
channel remain unaffected. In principle, the coupling constants
must be determined by matrix elements of Coulomb potential that
depend on details of wavefunctions and on chain arrangement, and
must contain an infinite sum over transversal reciprocal lattice
vectors. So expression (\ref{g2}) is not universal and depends on
material.

Eq. (\ref{g2}) leads to ${\mathbf q}$-dependent velocities
\begin{equation}
\omega_\rho = \frac{v_F}{K_\rho}q_\|, \quad \frac{1}{K_\rho} =
\sqrt{1 + \frac{\aleph}{s^2(q_\|^2 + \epsilon_\perp q_\perp^2)}},
\quad \aleph = \kappa^2s^2 = \frac{8e^2}{\hbar v_F},\label{C}
\end{equation}
where $\kappa$ is the inverse Thomas-Fermi screening length.

We do not perform explicit calculations restricting ourselves to
estimations. For the case of ${\mathbf q}$-dependent coupling
expressions for the thermodynamic potential and Green's functions
contain various integrals of correlation functions over ${\mathbf
q_\perp}$. One can show that the results obtained above can be
generalized qualitatively to the case of long-range Coulomb
interaction if we substitute $q_\perp$ in Eq. (\ref{C}) for its
characteristic value, $q_\perp \sim \pi/s$. For example, integrals
for corrections to the thermodynamic potential are dominated by
close values of coordinates and times, similar to Eq. (\ref{ln}),
and coupling parameters should be substituted by their averages
over $q_\perp$,
$$
\delta  = \frac{1}{4}(\overline{K_\rho} + \overline{1/K_\rho} +
K_\sigma + 1/K_\sigma - 4) \sim  \frac{1}{4} \left[\int \frac{s^2
d^2 q_\perp}{(2\pi)^2} \left(K_\rho + \frac{1}{K_\rho} \right) -
2\right].
$$
Note that $\aleph = 8\alpha (c/v_F)$, where $\alpha$ is the fine
structure constant. Since $v_F$ is much smaller than the velocity
of light, the factor $\aleph$ is large. For $v_F \approx 2 \times
10^7$ cm/s, which is typical value for transition metal
trihalcogenides, $\aleph \sim 90$. This corresponds to the case of
strong interaction and leads to quite large values of coupling
parameters, $\overline{1/K_\rho}  \sim
\sqrt{\aleph/\epsilon_\perp} \sim 3 \div 8,$ $\delta  \sim 1/2
\div 2.$

Now we discuss conditions for observation of the Luttinger liquid
in quasi-1D conductors stabilized by impurities. First we discuss
the condition for the temperature limiting from above the region
where the Luttinger liquid can exist. This condition reads, $T \ll
\omega_{1,\nu}$. The minimal excitation energies,
$\omega_{1,\nu}$, can be estimated as
$$\omega_{1,\nu} = \frac{\hbar v_F}{l}\overline{1/K_\nu}
\sim \varepsilon_F \left(\frac{1}{k_F l}\right) \overline{1/K_\nu}
\sim \varepsilon_F \, c_i \, \overline{1/K_\nu},$$ where $c_i$
stands for dimensionless impurity concentration corresponding to
number of impurities per one electron. As Fermi energy in NbSe$_3$
and TaS$_3$ is about 1 eV, we obtain that $\omega_{1,\rho}$ is
about 100 K for impurity concentration $c_i \sim 10^{-2} \div
10^{-3}$.

Another condition to be fulfilled is smallness of corrections to
the Green's function due to interchain hopping. According to
Eq.(\ref{G2e}) the corrections are small provided
$$
\left(\frac{t_\perp}{\varepsilon_F}\right)^2 c_i^{2\delta-2} \ll
1.
$$
If the interaction is strong enough, $\delta \gtrsim 1$, this
condition is not more strict than the condition for the limiting
temperature discussed above. For lower strength of interaction,
$\delta < 1$, this condition reads
$$
c_i \gg
\left(\frac{t_\perp}{\varepsilon_F}\right)^{1/(2-2\delta)}.
$$
Estimating $t_\perp$ as being of the order of the Peierls
transition temperature, $T_P \sim 100 \div 200$ K $\sim 0.01
\varepsilon_F $, we find that this condition can be fulfilled
easily even at $\delta = 1/2$ for relatively small impurity
concentration, $c_i \gg 10^{-2}$.

Thus we find that Luttinger liquid can be stabilized by impurities
in relatively pure linear-chain compounds at rather high
temperatures corresponding to experimental observation
Refs.\cite{ZZPM,ZZ,ZZGSZ} of the transition from metallic to
non-metallic conduction characterized by power law dependencies of
conductivity and by conductivity resembling the variable-range
hopping. However, in order to make detailed comparison with the
experimental data, calculation of the conductivity in a random
network made of weakly coupled bounded Luttinger liquids is
needed.

The author is grateful to S.V. Zaitsev-Zotov for useful
discussions. The work was supported by Russian Foundation for
Basic Research, by INTAS, CRDF, and by collaborative program with
the Netherlands Organization for Scientific Research.

\end{document}